\documentstyle[floats,epsf,prb,twocolumn,aps]{revtex}

\begin{document}

\draft

\preprint{...}
\wideabs{

\title{Phase separation in t-J ladders}

\author{Stefan Rommer and Steven R.\ White}

\address{Department of Physics and Astronomy, University of California,
Irvine, California 92697}

\author{D.\ J.\ Scalapino}

\address{Department of Physics, University of California,
Santa Barbara, California 93106}


\maketitle

\begin{abstract}

The phase separation boundary of isotropic t-J ladders is analyzed using
density matrix renormalization group techniques. The complete boundary to
phase  separation as a function of $J/t$ and doping is determined for a
chain and for ladders with two, three and four legs. Six-chain ladders have
been analyzed at low hole doping. We use a direct approach in which the
phase separation boundary is determined by measuring the hole density in 
the part of the system which contains both electrons and holes. In addition
we examine the binding energy of multi-hole clusters. An extrapolation in 
the number of legs suggests that the lowest $J/t$ for phase separation to
occur in the two dimensional t-J model is $J/t \sim 1$. 

\end{abstract}


} 


\narrowtext

\section{Introduction}

Phase separation in strongly correlated electron systems doped with holes
and its connection to high temperature superconductivity has attracted much
attention in the last ten years. 
Experimental evidence for phase separation\cite{jorgensen}  found in
${\rm La}_2{\rm CuO}_{4+\delta}$ and the desire to understand the full phase
diagram of strongly correlated models have led to intensive studies of this
effect in the t-J and Hubbard models. The
t-J model is one of the most basic models describing mobile
holes in an antiferromagnetic background. Nevertheless there is currently no 
general agreement for the phase separation (PS) boundary of the two
dimensional (2D) t-J 
model. At large interaction strengths $J/t$ the segregation of the system
into an antiferromagnetic region without holes and a hole-rich region is
well established. What happens in the physically relevant region for
smaller $J/t$ values and near half filling is, however, controversial.  

Emery, Kivelson and Lin suggested that the 2D t-J model phase separates at
all interaction strengths close to half filling based on variational
arguments and exact diagonalization of $4 \times 4$ clusters.\cite{emery} 
Later Putikka, Luchini and Rice found phase separation only for $J/t > 1.2$
using a high temperature expansion.\cite{putikka} Dagotto {\it et al.}
diagonalized $4 \times 4$ clusters and came to a similar
conclusion.\cite{dagotto2} They argued that at small $J/t$ there is a region
where holes bind into pairs but that PS starts only at $ J/t \sim
1$. Variational calculations and numerical studies on larger lattices 
by other authors have also yielded similar
results.\cite{valenti,prelovsek,fehske,kohno,poilblanc}

More recently Hellberg and Manousakis used a Green's function Monte Carlo
technique and found results in agreement with Emery {\it et al.}, i.e.\ PS
at all interaction strengths close to 
half filling.\cite{hellberg1}  On the other hand, subsequently Calandra,
Becca and Sorella used Green's function Monte Carlo and found evidence for
PS only for $J/t > 0.4$.\cite{calandra} White and Scalapino concluded in a
comparison of density matrix renormalization group (DMRG) results and other
numerical results by different authors that the 
ground state at small $J/t$ close to half filling should be striped and not
phase separated.\cite{white_compare} 

In this paper we use DMRG\cite{white} techniques to numerically study t-J 
ladders with up to six chains and determine the phase separation boundary as
a function of interaction strength and doping.  
This paper approaches the problem from a different direction than previous
studies. DMRG is highly accurate for narrow ladder systems, so here we
determine precise phase boundaries for narrow ladders and then consider what
happens as the ladders are made wider. As we will show in the next
section we find that an extrapolation in the width of the ladder systems
favors a lowest interaction strength of $J/t \sim 1$ for phase
separation to occur in a 2D system.  

Apart from being important to study the role of dimensionality on physical
properties and to give insight into 2D systems, ladder systems are very 
interesting in their own right, since they can be realized in various
materials. For example, it has been found that ${\rm La}_{1-x}{\rm Sr}_x{\rm 
CuO}_{2.5}$ consists of two chain ladders\cite{hiroinature} while ${\rm 
Sr}_{2n-2}{\rm Cu}_{2n}{\rm O}_{4n-2}$ has $n$-leg ladders weakly coupled to
each other.\cite{hiroi} Another example of a multiladder compound is ${\rm
La}_{4+4n}{\rm Cu}_{8+2n}{\rm O}_{14+8n}$.\cite{cava} 

The rest of this paper is organized as follows. In Section~\ref{mainresult},
before going into any details on our numerics and results, we present our
main findings for the phase separation boundary. In Section~\ref{model}
we present the model and the methods we have used. Section~\ref{results}
contains the detailed results and in Sec.~\ref{finalsec} we summarize and
conclude.     

\section{Phase Boundary} \label{mainresult}

\begin{figure}[t]
\epsfxsize=7cm
\centerline{\epsfbox{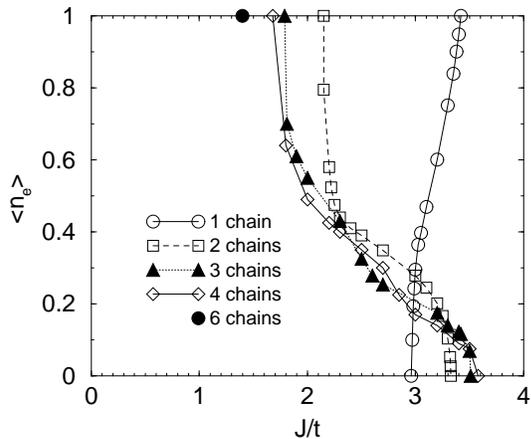}}
\caption{Boundary to phase separated region where $\langle n_e \rangle$ is
the total electron density of the system. Phase separation is realized to
the right of the curves.} 
\label{jc}
\end{figure}

Leaving the details to a later section we here present the main result. The
phase boundary for one, two, three and four leg ladders is shown in
Fig.~\ref{jc}. For the six chain ladders we have determined the PS
boundary at low hole doping as is indicated in the figure. A detailed
discussion of the method we have used to obtain the phase separation
boundary and the results for each ladder is given in Sec.~\ref{results}.

The DMRG has proven to be an accurate
method for studying one dimensional systems and ladder systems of
limited width. The results presented in Fig.~\ref{jc} and in
Sec.~\ref{results} for systems with up to six legs 
should be very reliable and thus settle the issue of the phase boundary
for these ladders. What is less clear is whether, knowing that the result
for each ladder is accurate, it is possible to find the behavior of a
2D system by doing an extrapolation in the number of legs. 

\begin{figure}[t]
\epsfxsize=7cm
\centerline{\epsfbox{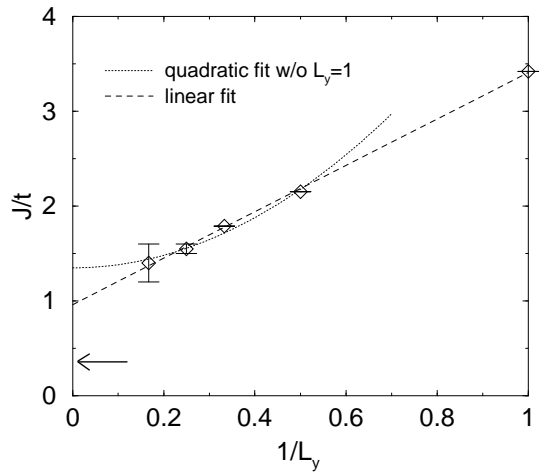}}
\caption{Phase separation boundary at low hole doping as a function of the
inverse number of legs. The arrow indicates $J/t \approx 0.35$, a typical
value of $J/t$ for the cuprates.} 
\label{jc_ext}
\end{figure}

Since the controversy about 2D systems has primarily been about the PS boundary
close to half filling, we will focus on the critical $J/t$ value for phase
separation for $\langle n_e \rangle = 1$, where $\langle n_e \rangle$ is the
average electron density.  We can get an estimate for the
2D model by extrapolating our data in the number of legs as shown in
Fig.~\ref{jc_ext}. We find the  PS boundary at about $J/t \sim 1$. From our
data we cannot tell the exact functional form of the convergence to
2D. However, as can be seen in Fig.~\ref{jc_ext}, the qualitative conclusion
made above does not depend on whether the convergence is linear or quadratic
in the inverse number of legs. 

\section{Model and Method} \label{model}

The Hamiltonian of the t-J model is
\begin{equation}
H = -t \sum_{\langle i,j \rangle, s} \left( c_{i s}^\dagger c_{j
s} + {\rm H.c.} \right) + J \sum_{\langle i,j \rangle} \left( {\bf S}_i
\cdot {\bf S}_j - \frac{n_i n_j}{4} \right) ,
\end{equation}
with the constraint of no doubly occupied sites and where $\langle i,j
\rangle$ denotes nearest-neighbor sites, $s (= \uparrow {\rm or} 
\downarrow)$ is a spin index, $c_{is}^\dagger$ ($c_{is}$) is the 
electron creation (annihilation) operator for an electron at site $i$ with
spin $s$, ${\bf S}_i = c_{i s}^\dagger {\bf \sigma}_{s,s^\prime} c_{i
s^\prime}$ is the spin-1/2 operator and $n_i = \sum_s  c_{is}^\dagger
c_{is}$. The nearest-neighbor hopping interaction is $t$ and the
nearest-neighbor exchange interaction is $J$. We consider ladders with equal  
couplings along the legs and the rungs. 


At half filling the t-J model is equivalent to the Heisenberg model. When
holes are added to the system antiferromagnetic links are removed and this
increases the energy.  To reduce the number of broken antiferromagnetic
bonds the holes prefer to stay close together and this can result in a phase
separated system. The kinetic part of the Hamiltonian would, however, prefer
the holes to spread out and this competition between the kinetic and
exchange energy results in a rich phase diagram.   

In a t-J ladder with two or more legs, the phase separated system consists of 
one region containing holes and one region without any holes.  For a phase
separated system with a given  
$J/t$ and hole doping $x$, the density of holes in the hole-rich region 
gives us a point on the phase boundary curve separating phase separated and
uniform systems. Using a Maxwell construction, this
can easily be realized by examining the energy per site of the lowest energy 
{\em uniform} state as a function of the hole doping, $e(x)$, where $x$ is
the total hole doping of the system. Without any
holes this energy will be that of the Heisenberg model $e(x=0) = 
e_H$. Now consider a straight line connecting $e(x)$ between the Heisenberg
point, $x=0$, and some finite doping, $x_{\rm ps}$,
\begin{equation}
e_{\rm lin}(x) = e_H + x \frac{e(x_{\rm ps}) - e_H}{x_{\rm ps}} . \label{elin}
\end{equation}
If at some doping $x$, where $0 < x < x_{\rm ps}$, we have  
\begin{equation}
e_{\rm lin}(x) < e(x) \label{xc}
\end{equation}
it is possible for the uniform system to lower its energy by forming two
separate phases, one with hole density $x_{\rm ps}$ and one with no
holes. The inequality~(\ref{xc}) will be fulfilled when $e(x)$ has a
negative second derivative. In this case the density of the hole-rich
region, $x_{\rm ps}$, is found by  minimizing Eq.~(\ref{elin}) with respect
to $x_{\rm ps}$. It is easily shown that this corresponds to an $x_{\rm ps}$
such that the slope of $e(x)$ at $x_{\rm ps}$ is equal to the slope of the
straight line $e_{\rm lin}(x)$.\cite{emery} Since this is valid for all
dopings $0 < x < x_{\rm ps}$, the density $x_{\rm ps}$ is thus independent of
$x$ and equal to the value of the hole 
density at which phase separation sets in. The energy per site of the phase
separated state will be linear in the doping and will be given by $e_{\rm
lin}(x)$. For large enough value of $J/t$, we have a fully phase separated
system with  $x_{\rm ps}=1$. In this case the system consists of a
Heisenberg $(x=0)$ region and a completely electron-free $(x=1)$ region.

For a single chain at large enough $J/t$ the situation is the same: the
system is fully phase separated and has a Heisenberg phase and an empty
phase. However, at intermediate values of $J/t$ the chain and the ladders
are quite different. If we reduce $J/t$ for the ladders or for a two
dimensional system, the empty phase becomes unstable first 
and electron pairs evaporate into the electron-free region.\cite{emery}
However, as we reduce $J/t$ in a single chain it is the Heisenberg phase that
gets destroyed first and holes evaporate into this hole-free
region.\cite{ogata}  This means that for an intermediate
range of $J/t$ the chain has one region without any electrons and one region
with electrons and holes. This is seen in Fig.~\ref{jc} where the PS boundary
for a chain occurs at a smaller $J/t$ at low electron density than at low
hole doping while the opposite is true for the ladders. 

For chains as with
the ladders, the density of holes in the mixed phase (containing both
electrons and holes) is equal to the boundary density at which phase
separation sets in. 
Thus, by determining the hole density in the hole rich region we
automatically know the PS boundary.  The problem of mapping out the boundary  
line in the two dimensional $(J/t,x)$ space is then reduced to a one
dimensional problem: for each $J/t$ we get one point on the PS boundary (as
long as the system is phase separated). 
Using DMRG we can determine the PS boundary in a very direct way by
actually measuring the hole density in the mixed phase.

The initial positions of the holes are determined during the DMRG 
system buildup sweep by choosing appropriate quantum numbers for the target
state  in each iteration. After this sweep a number of finite size sweeps
are done. 
Starting with a uniform hole distribution and letting a possible phase
separated state form during the finite size sweeps is a possible but rather
slowly converging method. It is better to distribute the holes uniformly in
only part of the system and let the DMRG converge from there. By doing
several runs for the same hole doping and $J/t$, each time adjusting the
initial hole configuration according to the
previous results in a self consistent fashion, it is possible to speed up
convergence. When the DMRG has converged we measure the total hole density
in the rungs,  
\begin{equation}
\langle n_r(l_x) \rangle = \sum_{l_y} 1 - \langle c_{l_x,l_y}^\dagger
c_{l_x,l_y}  \rangle , \label{nrung} 
\end{equation}
and determine $x_{\rm ps}$. Here $c_{l_x,l_y}$ is the electron annihilation
operator at site $(l_x,l_y)$.

At very low hole doping it becomes more difficult for the DMRG to converge
to  a uniform density of holes in the hole-rich phase.  
To estimate the PS boundary at low hole doping we therefore also consider
the binding of holes. At intermediate couplings $J/t$ the holes bind 
to form many-hole bound states (pairs and domain-walls) as will be discussed
in more detail for each ladder system 
below. The binding energy of two domain-walls (stripes) with $h$ holes each
is defined by 
\begin{equation}
E_b(h) = 2 \left( E_h - E_0 \right) - \left( E_{2h} - E_0 \right) ,
\label{ebind} 
\end{equation}
where $E_h$ is the ground state energy of a system with $h$ holes. For
couplings such that a  $2h$-hole compound breaks up into two $h$-hole
stripes the  binding energy is zero. With only a few holes in a bound state
located around the center of the ladder we have been able to use long enough 
ladders so that the holes are always far from the open boundaries at the
ends of the legs. The binding energy in Eq.~(\ref{ebind}) is then
nearly independent of the system length. By determining the coupling
$J/t$ at which the $2h$-hole complex breaks apart, i.e.\ where $E_b = 0$, we
find an estimate for the PS boundary in the limit of low hole doping. 

There is of course the possibility that the binding of two stripes does not 
signal phase separation but instead corresponds to the formation of another
state with twice as many holes in each stripe. By comparing the estimates
for the PS 
boundary for finite hole density determined by measuring the density in the
hole-rich phase as described above with estimates from the binding of
stripes we find that the two results agree well at low hole doping. We thus
believe that this does not occur.   
Another exotic possibility is that a striped state is not uniformly striped
but actually phase separated into a striped phase with a fixed, fairly large
spacing between stripes and a Heisenberg
phase. We have not seen indications for such a phase in our
calculations. However, if this were to occur at very low hole doping (say $x
\sim 0.01$) we would not be able to detect it, particularly on the wider
ladders. 


We believe that our method of calculating the PS boundary value $J/t$ is a
more accurate method than estimating $ x_{\rm ps} $ by a minimization 
of $ e(x)_{\rm lin} $, especially at low hole doping where the minimum
occurs very close to the Heisenberg point $x=0$.  

Since the DMRG is most accurate for systems with open boundary conditions we
have used open boundary conditions (BC) in the leg direction. We have
however performed calculations on several system lengths to examine the
effects of the open edges. For the ladders with four legs or less we
were able to treat long enough systems to make finite length effects
negligible. For the six chain ladder we were more limited in length but we
still believe our results to be accurate.

For the two, three and four chain ladders we have used open boundary
conditions also in the rung direction. It has been found that three
and four hole stripes form in the rung direction on the open three and four
leg ladders respectively.\cite{white3chain,white4chain} Macroscopic phase
separation, which can be viewed as the attraction and binding of stripes,
can thus be naturally studied in these systems. We have further found that
stripes in the rung direction form in six leg ladders with cylindrical
boundary conditions, i.e.\ 
periodic BC in the rung direction. On the other hand, in open six chain
ladders with reasonably large $J/t$ the holes tend to localize on the outer
legs leaving the middle legs hole free, at least partly because only three
antiferromagnetic bonds will be broken if a hole is positioned at an open
boundary compared to four broken bonds for a hole in the bulk and the gain
in exchange energy is larger than the increase in kinetic energy due to
localization at the boundaries. Therefore we used cylindrical BC with all
six chain ladders studied in this article. For the three and four leg
systems near and in the phase separated region we always found many-hole
complexes that were essentially uniform in the rung direction even with open
BC.   

We keep at most 2400 states in our DMRG calculations and the truncation errors
range from less than $10^{-7}$ for the single chain to less than $10^{-4}$
for the six-chain ladder. To reduce effects from the open boundaries at the
ends of the legs we sometimes apply a small local chemical potential at the
outermost rungs. This reduces the attraction of holes to the edges and makes
the hole density close to the edges smoother. A local chemical potential
between $\mu \approx 0.2(J/t)$ and $\mu \approx 0.5(J/t)$ was typically used.

\section{Results} \label{results}

\subsection{One chain}

\begin{figure}[t]
\epsfxsize=7cm
\centerline{\epsfbox{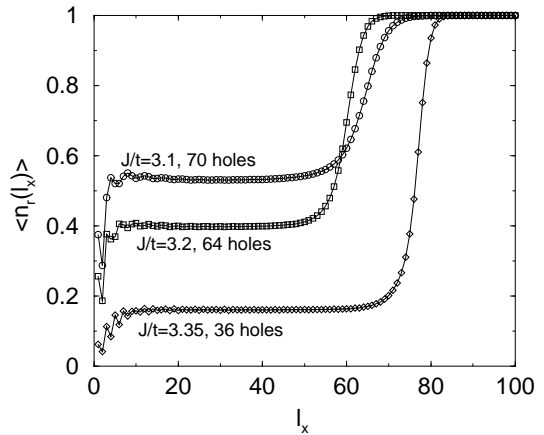}}
\caption{Density of holes along a $100$-site chain for three
values of $J/t$.} 
\label{hrung1chain}
\end{figure}

We first consider the simplest case, a single chain with open boundary
conditions. 
The hole density along the chain for three values of $J/t$ is shown in
Fig.~\ref{hrung1chain}. As can be seen, the system separates into an
electron-free region and a region with both holes and electrons. 
By measuring the hole density in the mixed region for a given $J/t$ we
directly find the phase separation boundary $x_{\rm ps}$ as discussed in the
previous section.  The complete phase separation boundary is plotted in
Fig.~\ref{jc}. Our results agree well with the results at intermediate
doping by Ogata {\em et al.},\cite{ogata} but they did not determine the
phase boundary at very low hole doping or very low electron densities. We
also find reasonable agreement with the results of Hellberg and
Mele\cite{hellberg1chain} although there are small differences at low and
high dopings. However, we believe our results for a single chain are more
accurate than previous studies.

\subsection{Two chains}

\begin{figure}[t]
\epsfxsize=7cm
\centerline{\epsfbox{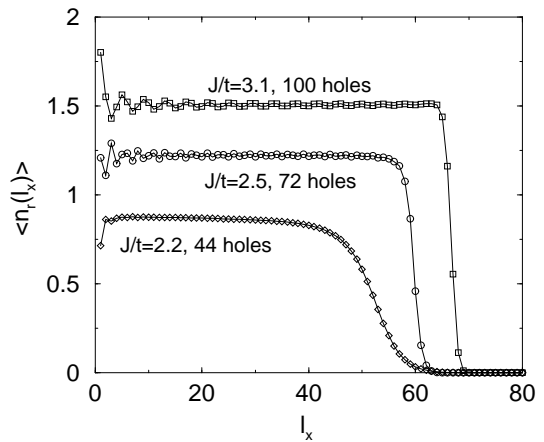}}
\caption{Density of holes in the rungs for an $80 \times 2$
system.}  
\label{ps2chain}
\end{figure}

We next consider a two-chain ladder with open boundary conditions. 
Some of the data for the two chain systems shown here were originally
presented in Ref.\ \onlinecite{sierra}, where it was found that a recurrent
variational ansatz gave qualitatively similar results. 
Here we present additional results and explain the calculational technique
in more detail.
We have calculated the hole density for $80 \times 2$ systems at different
coupling parameters $J/t$. The total hole density in the rungs, as defined 
in Eq.~(\ref{nrung}),  is shown in Fig.~\ref{ps2chain} for three values of
$J/t$. As seen the system now divides into a hole-free region and a
mixed region.

\begin{figure}[t]
\epsfxsize=7cm
\centerline{\epsfbox{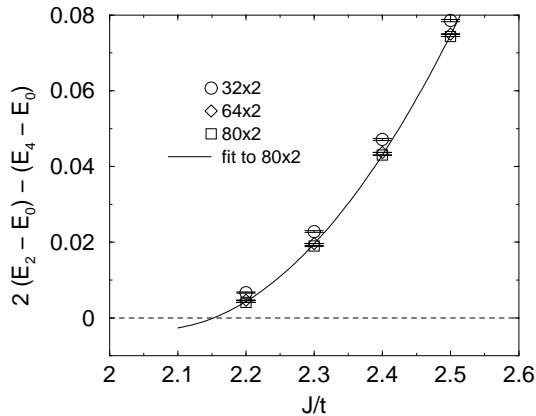}}
\caption{Binding energy of two pairs in $32 \times 2$, $64 \times 2$ and $80
\times 2$ 
ladders. The fit in the figure to the $80 \times 2$ data is $1.751-1.699
J/t+0.4114 (J/t)^2$ which gives zero binding energy at $J/t \approx
2.156$. } 
\label{bind2chain}
\end{figure}

For $J/t$ just smaller than where the model phase separates we
find that four-hole clusters split up into two pairs but that pairs are
stable. Inside the phase separated phase the four-hole clusters of course
are stable and we have determined the binding energy of 
two pairs, $E_b(2)$, as defined in Eq.~(\ref{ebind}). The
binding energies of two pairs on $32 \times 2$, $64 \times 2$ and $80 \times
2$  lattices are shown in Fig.~\ref{bind2chain}. The extrapolated value at
zero binding energy for the $80 \times 2$ system is $J/t \approx 2.15$.
With only two or four holes located close to the center of a long lattice
the energy per hole should hardly be affected by the finite size. The
calculations on the different lattices in Fig.~\ref{bind2chain} support this.  
The value of $J/t$ where the binding energy becomes zero is thus an estimate
of the PS boundary in the limit of zero hole doping, assuming that there is no
intermediate phase where four holes bind but the system does not phase
separate. The boundary value $J/t \approx 2.15$ does however connect
smoothly with the boundary at larger hole doping (see Fig.~\ref{jc}) and we 
therefore exclude the possibility of a non phase separated phase with
four-hole bound states.

Our results for the PS boundary agree well at low hole doping with the
results of Refs.~\onlinecite{tsunetsugu} and \onlinecite{hayward} although
they find a slightly larger $J/t$ at larger dopings. This
might be attributed to their method of using the divergence of the
compressibility as a signature of the onset of phase 
separation. This criterion has been shown to give a  slightly  larger value
of $J/t$ at the PS boundary for finite systems.\cite{hellberg1} 
Again, for the two chain case we believe our results are the most accurate
to date. 

\subsection{Three chains}

\begin{figure}[t]
\epsfxsize=7cm
\centerline{\epsfbox{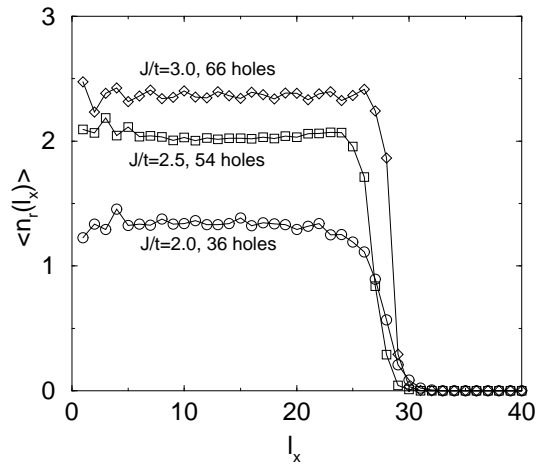}}
\caption{Density of holes in the rungs of a $40 \times 3$ system for three
values of $J/t$.} 
\label{hrung3chain}
\end{figure}

We now consider a three chain ladder with open boundary conditions. 
We have determined the phase separation boundary by calculating the hole
density of a $40 \times 3$ system at different dopings. The hole density in
the rungs of three such systems is shown in Fig.~\ref{hrung3chain}. 

The ground state properties of this system at $J/t=0.35$ has been studied by
White and Scalapino.\cite{white3chain} It was found that at low hole doping
the holes form a dilute gas of uniform density. At higher doping the holes
were found to bind into three-hole domain walls. 

\begin{figure}[t]
\epsfxsize=7cm
\centerline{\epsfbox{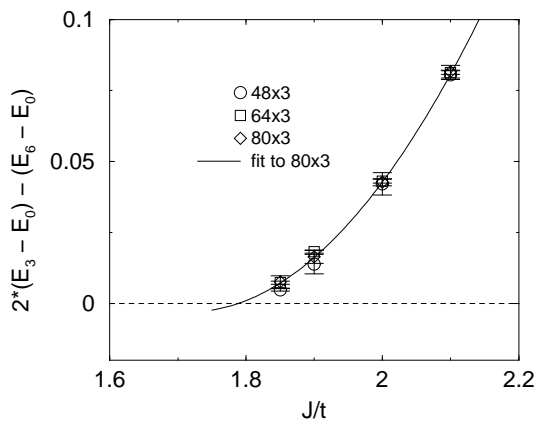}}
\caption{Binding energy of two three-hole stripes in $48
\times 3$, $64 \times 3$ and $80 \times 3$ systems. The fit in the figure to
the $80 \times 3$ data is $1.676 - 1.956 J/t + 0.5697 (J/t)^2$ which crosses
zero binding energy at $J/t \approx 1.79$.}   
\label{bind3chain}
\end{figure}

We find that just outside the PS boundary a six-hole cluster splits up into
two stable three-hole domain walls. To make an estimate of the PS boundary
at low hole doping we calculated the binding energy of two three-hole domain
walls. Figure~\ref{bind3chain} shows the binding energy for $32 \times 3$,
$48 \times 3$ and $80 \times 3$ systems.  The extrapolated value at
zero binding energy for the $80 \times 3$ lattice, $J \approx 1.79$,
connects smoothly with the estimates at  larger hole density. Many-hole
bound states with more than three holes were thus not considered. The
complete PS boundary line is plotted in Fig.~\ref{jc}.

\subsection{Four chains}

\begin{figure}[t]
\epsfxsize=7cm
\centerline{\epsfbox{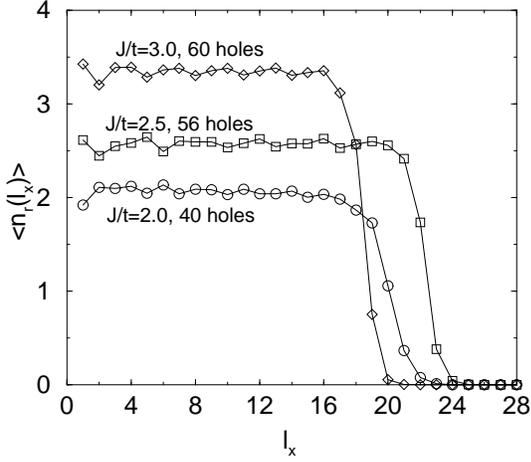}}
\caption{Density of holes in the rungs of a $28 \times 4$ system for three
values of $J/t$.} 
\label{hrung4chain}
\end{figure}

We next consider a four chain ladder with open boundary conditions. 
We have determined the phase separation boundary by measuring the hole
density in the hole-rich phase of a phase separated $28 \times 4$
system. The density in the rungs in three systems is shown in
Fig.~\ref{hrung4chain}. 

\begin{figure}[t]
\epsfxsize=7cm
\centerline{\epsfbox{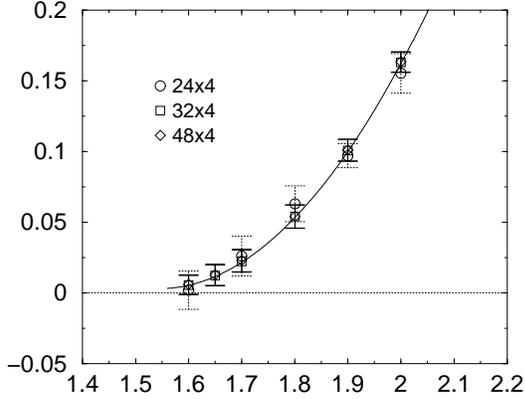}}
\caption{Binding energy of two four-hole stripes in $24 \times 4$, $32
\times 4$ and $48 \times 4$ systems. The binding energy becomes zero at $J/t
\approx 1.55 \pm 0.05$.}  
\label{bind4chain}
\end{figure}

White and Scalapino have found that, depending on $J/t$ and the hole doping,
the ground state of this system is a state containing a dilute gas of hole
pairs, a striped charge density wave domain-wall state or a phase separated
state.\cite{white4chain} We find that just outside the PS region an eight-hole
cluster is unstable and splits up into two stable four-hole domain walls. We
have calculated the binding energy of two four-hole domain walls inside the
PS region on $24 \times 4$, $32 \times 4$ and $48 \times 4$ lattices. The
result is shown in Fig.~\ref{bind4chain}. The extrapolated value at
zero binding energy, $J/t \approx 1.55 \pm 0.05$, is (again) an estimate for
the phase separation boundary in the limit of small hole doping. Non
phase-separated states with bound states of more than four holes were not 
considered.  The whole phase separation boundary for four chains is shown in
Fig.~\ref{jc}.

\begin{figure}[t]
\epsfxsize=7cm
\centerline{\epsfbox{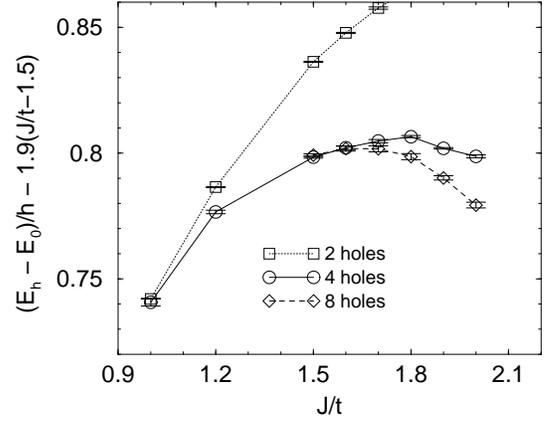}}
\caption{Energy per hole in a $24 \times 4$ system. Error bars are of the
same size as the symbols or smaller. Note that the energies have been
shifted by $-1.9(J/t-1.5)$ to increase clarity of the figure. Phase
separation occurs for $J/t \protect\agt 1.55$.} 
\label{eperhole4chain}
\end{figure}

As a complement to Fig.~\ref{bind4chain} the energy per hole of different
size clusters is shown in Fig.~\ref{eperhole4chain}. For $J/t$ slightly
smaller than $1.6$ the eight-hole cluster splits up into two four-hole
clusters that, when well separated, become degenerate in the energy per hole
with a single four-hole cluster. As can be seen the four-hole bound states
are preferred over pairs just outside the PS region.  

\subsection{Six chains}

\begin{figure}[t]
\epsfxsize=7cm
\centerline{\epsfbox{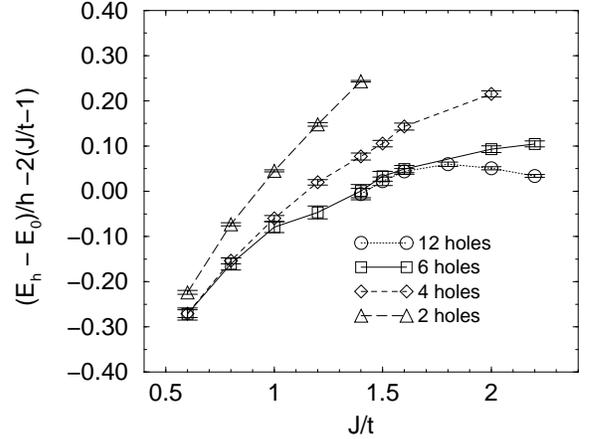}}
\caption{Energy per hole in a $12 \times 6$ system. Error bars are of the
same size as the symbols or smaller. Note that the energy per hole has been 
shifted by $-2(J/t-1)$ to increase clarity of the figure.} 
\label{eperhole6chain}
\end{figure}

We next consider a six chain ladder with periodic boundary conditions in
the rung direction. 
Here, we find that at low hole doping and in the range of $J/t$ studied, $0.6 \leq
$J/t$ \leq 2.3$,  the ground state of this system was either
a state containing domain walls with four holes, domain walls with six
holes or a phase separated state. The energy per hole in a $12 \times 6$
system with different numbers of holes located around the central rungs is
shown in Fig.~\ref{eperhole6chain}. We believe that the region were a twelve 
hole compound has lower energy per hole than a six hole compound is phase
separated. There is thus a region with six hole domain walls between the
phase separated region and the region with four hole domain walls. 

\begin{figure}[t]
\epsfxsize=7cm
\centerline{\epsfbox{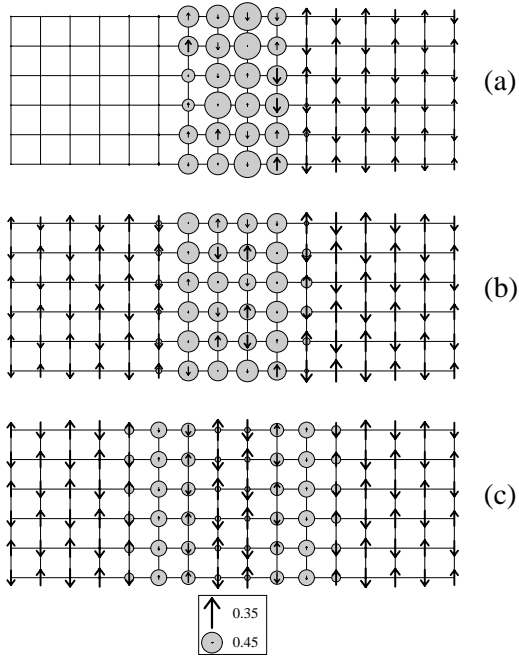}}
\caption{Sweeps 2 (a), 8 (b) and 23 (c) with 12 holes at $J/t = 1$ on a $16
\times 6$ lattice. The energies are  $-118.1$, $-120.2$ and $-122.1$
respectively. The diameter of the shaded circles and the lengths of the
arrows indicate the hole density and the expectation value of $S^z$ 
respectively. The initial configuration has all holes in the center of the
system. As the DMRG sweeps through the system the total energy is reduced as
the twelve holes split into two six-hole stripes.}   
\label{hsz6chain}
\end{figure}

In the intermediate region $0.9 \alt J/t \alt 1.4$, where the six hole domain
has the lowest energy per hole, a twelve hole compound should be unstable
and break up into two 
six-hole compounds. In Fig.~\ref{hsz6chain} the hole density in three
intermediate finite size sweeps from one DMRG calculation with $J/t=1$ is
shown. As the 
DMRG sweeps through the lattice, local changes in the approximation of the
ground state wave function are done to minimize its energy. In the first
sweep all holes were located in the center of the system but to reduce the
energy they split into two six-hole stripes. A similar run with $J/t=0.6$
results in a split of the twelve holes into three four-hole 
stripes (not shown) in agreement with Fig.~\ref{eperhole6chain}.

\begin{figure}[t]
\epsfxsize=7cm
\centerline{\epsfbox{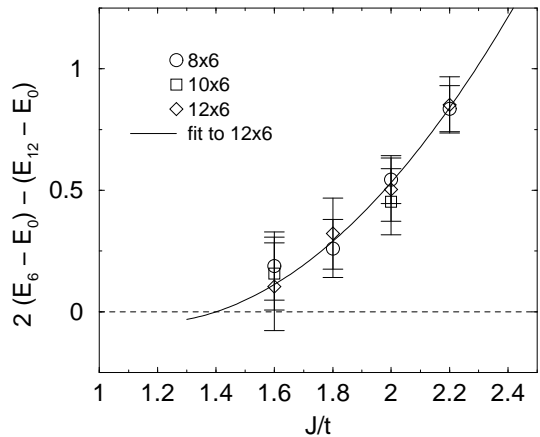}}
\caption{Binding energy of two six-hole stripes in $8 \times 6$, $10
\times 6$ and $12 \times 6$ systems. 
The binding energy becomes zero at $J/t \approx 1.4 \pm 0.2$.} 
\label{bind6chain}
\end{figure}

To estimate the phase separation boundary at low hole doping we have
calculated the binding energy of two six-holes stripes in $8 \times 6$, $10
\times 6$ and $12 \times 6$
systems as shown in Fig.~\ref{bind6chain}. The extrapolated value at which
the binding energy crosses zero is $J/t \approx 1.4$. This data point is
also shown in the phase diagram in Fig.~\ref{jc}.

\section{Summary} \label{finalsec}

In this paper we have studied phase separation in t-J ladders with one, two,
three, four and six legs. We have mapped out the complete phase separation
boundary as a function of interaction strength $J/t$ and doping. We have
used a direct approach by actually calculating the density of holes in the
mixed phase. At very low hole doping 
there are few holes in the system and this approach becomes
difficult. In this case we have estimated the phase separation boundary by
calculating the binding energy of stable many-hole complexes. In this case,
extrapolation to zero binding energy gives us the estimate for the PS
boundary.  The critical $J/t$ for half filling and zero electron density are
shown in Table~\ref{jctab}.

\begin{table}
\begin{tabular}{llll}
$L_y$ 	& BC & $J_c/t$ for $n_e=1$ & $J_c/t$ for $n_e=0$ \\ \hline
1 & OBC		& 3.415(2) & 2.96(1) \\
2 & OBC 	& 2.156(2) & 3.327(1) \\
3 & OBC 	& 1.79(2)  & 3.51(1) \\
4 & OBC 	& 1.55(5)  & 3.58(1) \\
6 & PBC 	& 1.4(2)   & 3.25(5) \\ 
\end{tabular}
\caption{Critical $J/t$ for half filling ($n_e=1$) and zero electron density
($n_e=0$). The column $L_y$ is the number of legs and BC denotes whether
periodic or open boundary conditions in the rung direction were used.}
\label{jctab}
\end{table}


To investigate the possibility of a lowest limit on $J/t$ for phase
separation to occur in a 2D system we made an extrapolation in the number of
legs. This extrapolation suggests that the 2D t-J model only phase separates
for $J/t \agt 1$, in agreement with several previous studies.

\section*{Acknowledgements}

S.\ Rommer is thankful for support from the Swedish Foundation for 
International Cooperation in Research and Higher Education (STINT). S.R.\
White acknowledges support from the NSF under grant \#DMR98-70930 and D.J.\ 
Scalapino acknowledges support from the NSF under grant \#DMR98-17242.



\end{document}